\documentclass{ws-ijmpa}
\usepackage[super,compress]{cite}
\usepackage{graphicx}
\usepackage{soul}  
\usepackage{xcolor}

\begin{document}
\markboth{Hemida.H.Mohammed}{Instructions for typing manuscripts (paper's title)}

%
\catchline{}{}{}{}{}
%

\title{Effect of an Expanding Charged Cloud on
two-particle Bose-Einstein Correlations}

\author{ Hemida H. Mohammed$^{1,2}$,  M\'at\'e  Csan\'ad$^3$, Y. Mohammed$^{1,2}$,  N. Rashed$^1$, D\'aniel Kincses$^3$,    M. A. Mahmoud$^{1,2}$}

\address{$^1$ Physics Department, Fayoum University, El-Fayoum, 63514, Egypt\\ 
$^2$ Center for High Energy Physics (CHEP-FU), Fayoum University, El-Fayoum, 63514, Egypt\\
$^3$ ELTE E\"otv\"os Lor\'and University, P\'azm\'any P. s. 1/a, 1117 Budapest, Hungary}

\maketitle{}

\begin{history}
\received{Day Month Year}
\revised{Day Month Year}
\end{history}

\begin{abstract}
In high-energy physics, quantum statistical correlation measurements are very important for getting a good picture of how a particle-emitting source is structured in space and time, as well as its thermodynamic properties and inner dynamics. It is necessary to take into account the various final state effects since they have the potential to alter the observed femtoscopic correlation functions. Protons are affected mostly by the strong interaction, whereas other charged particles are mostly influenced by the Coulomb interaction. The interaction of the particles under investigation with the fireball or the expanding cloud of the other particles in the final state might also have significant consequences. This may cause the particle{\'{s}} trajectory to shift. This phenomenon can be viewed as an Aharonov-Bohm effect since the pair's alternate tracks reveal a closed loop with an internal field. We investigate a numerical solution for a toy model to study the modifications of Bose-Einstien correlation function strength, which is sensitive to this effect.

\end{abstract}

\ccode{PACS numbers:}


\section{Introduction}	
After hadronization in heavy-ion collisions, a high multiplicity of pions~\cite{lisa2009multiplicity}, kaons, and other charged particles are produced. The phenomenon of correlated particles can arise from several physical processes, e.g., jets, collective flow, conservation laws, and resonance decays. Another important source of correlation can be the Bose-Einstein~\cite{Csorgo:1994in} or Hanbury-Brown-Twiss (HBT)~\cite{Hanbry-Twiss:1956corr,Hanbry-Twiss:1956test} effect, where correlation functions are used to discover the geometry of the emitting source. In the case of two identically charged pions, these correlations are the major source of the momentum correlations owing to the indistinguishability of the two identical pions and their symmetrical pair wave functions. Technically, several experimental effects have to be considered that could modify the correlation functions~\cite{aggarwal2011pion}. The space-time geometry of the particle emitting source~\cite{lednicky2001femtoscopy,lisa2005femtoscopy,PHENIX:2017ino,Korodi:2022ohn,Csanad:2024hva} may be explored by measuring Bose-Einstein correlation functions of identical particles. In addition to quantum statistics, multiple phenomena affect the measured momentum correlations of the identical pair, the most significant being the final-state interactions~\cite{Kincses:2019rug,Nagy:2023zbg}.
A further modification to the momentum correlations is due to an important effect of the interaction between any given pair of identical hadrons and the surrounding charged particles (gas cloud of charged particles) due to the potential of the charged particles in the absence of the EM field which could be interpreted as an Aharonov--Bohm like effect~\cite{Aharonov,Csanad:2020qtu}. The pair wave function experiences a phase-shift proportionate to the flux encircled by the path due to the (electromagnetic, strong, etc.) fields inside this closed path ~\cite{Csorgo:1999sj,lednicky2011femtoscopic,Csanad:2020qtu}. The influence of the quantum-statistical correlations is modified by the phase shift, as will be covered later. We show in this work, how this phenomenon may be used in heavy-ion experiments.~\cite{Csanad:2020qtu}

\begin{figure}
\centering
\includegraphics[width=0.6\textwidth]{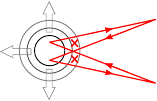}
\caption{\label{fig:ahbhubble} Aharonov--Bohm effect with a Hubble expanding source}
\end{figure}

This paper is structured as follows. It starts with an introduction about Bose-Einstein correlations in high-energy physics in Section~\ref{s:becintro}. In Section~\ref{s:lambda23} the role of randomly fluctuating fields in multiparticle correlations is discussed.
Then, in Section~\ref{s:toymodel} numerical calculations are set up with a toy model to give quantitative details on how such final-state effects affect the Bose-Einstein correlation functions. Finally, in Section \ref{s:results} an observable sensitive to this effect is presented, followed by conclusions.

\section{Bose-Einstein correlations in relativistic collisions}\label{s:becintro}

In general, the two-particle momentum correlation function is defined  as
\begin{align}
C_2(p_1,p_2)\equiv\frac{N_2(p_1,p_2)}{N_1(p_1)N_1(p_2)},\label{e:c2def}
\end{align}
where $N_2(p_1,p_2)$ and  $N_1(p)$ are the two-- and one-particle invariant momentum distributions, with $p_i$ being the
(four-)momenta of the bosons. The pair wave function of charged bosons is a plane wave by ignoring final-state interactions. This allows the correlation function to be expressed using the phase-space density of the particle-emitting source $S(x,p)$ as
\begin{equation}\label{e:Cp1p2:general}
C_2(p_1,p_2)= 1+ Re\frac{\widetilde S(q,p_1)\widetilde S^*(q,p_2)}{\widetilde S(0,p_1)\widetilde S^*(0,p_2)} ,
\end{equation}
where $q\equiv p_{1}-p_{2}$ is the relative momentum, complex conjugation is denoted by $^*$, and $\widetilde S(q,p)$ denotes
the Fourier transform of the source:
\begin{equation}\label{e:tildeSdef}
\widetilde S(q,p)\equiv\int S(x,p) e^{iqx} d^4x .
\end{equation}

For source functions and typical kinematic domains encountered in correlation measurements in heavy-ion collisions,
the dependence of $\widetilde S(q,p)$ as defined in Eq.~\eqref{e:tildeSdef} is much smoother in the original $p$ momentum variable than in the relative momentum $q$ coming from the Fourier transform, so it is customary to substitute $p_1$ and $p_2$ in Eq.~\eqref{e:Cp1p2:general} with an average value of $K\equiv \frac{1}{2}(p_1+p_2)$. Hence if
$p_1\approx p_2\approx K$, Eq.~\eqref{e:Cp1p2:general} gives the usual formula of
\begin{equation}\label{e:C2expr}
C_2(q,K)=1+\frac{|\widetilde S(q,K)|^2}{|\widetilde S(0,K)|^2}.
\end{equation}
Here the dependence on the pair momentum $K$ is smoother than on $q$, hence one usually thinks of $q$ as the ``main''
kinematic variable, and by some parametrization for the $q$ dependence, one explores the $K$ dependence of
the parameters of this parametric function. It is important to note furthermore, that while the above is only
true for non-interacting bosons, final state interactions such as the Coulomb interaction can be handled in a simple
and effective manner, outlined e.g. in Ref.~\cite{Sinyukov:1998fc,maj2009coulomb}. This usually involves the definition of a ``Coulomb-correction''
that removes the effect of the Coulomb interaction, and the corrected correlation functions can be analyzed then according to Eq.~\eqref{e:C2expr}.

From Eq.~\eqref{e:C2expr} it is clear that the correlation function takes the value $2$ at zero relative momentum.
However, experimentally $C_2(0) = 1+\lambda_2$, where $\lambda_2$ is
the so-called intercept parameter (or the strength of the correlation function),
and usually $\lambda_2 \le 1$ holds. The formula for the correlation function then may be empirically modified as
\begin{equation}\label{e:C2:corehalo}
C_2(q,K)=1+\lambda_2\frac{|\widetilde S(q,K)|^2}{|\widetilde S(0,K)|^2} .
\end{equation}
The reason for $\lambda_2<1$ can be coherent pion (or kaon) production~\cite{Bolz:1992hc}, but for pions, a simpler
explanation is provided by the core-halo model~\cite{Csorgo:1994in}. This treats the source as a sum of two components.
One is the \emph{core}: the primordial pion source, which is the main reaction zone produced in a heavy-ion collision,
whose Fourier transform is resolvable in momentum ($q$--)space by the correlation measurement. The
other component is a much wider \emph{halo}, the contribution of the pions that are decay products of long-lived
resonances (that travel much farther than $\simeq 10$ fm: $\eta$, $\eta'$, $\omega$, $K^0_S$, etc). The Fourier
transform of the broad halo would be a very sharp peak at $q=0$, and this is experimentally essentially
unresolvable. The intercept $\lambda_2$, i.e. the extrapolation of the measured \emph{visible} correlation function to zero
relative momentum) is then basically the square of the fraction $f_c$ of pions coming from the core (since both pions have
to come from the core if they are to contribute to the visible correlation function). For the details see
Ref.~\cite{Csorgo:1994in}. So in the core-halo picture
\begin{equation}\label{e:lambda:corehalo}
     \lambda_2 = f_c^2,\quad f_c = \frac{N_{core}}{N_{core}+N_{halo}} .
\end{equation}

Hence $\lambda_2$, the strength of the two-particle correlation measures
the fraction of primordial pions. This leads to an interesting application.
It is known~\cite{Kapusta:1995ww} that in case of chiral U$_A(1)$
symmetry restoration the mass of the $\eta'$ boson (the ninth, would-be
Goldstone boson) is decreased, thus its production cross section is
heavily enhanced. The $\eta'$ has a decay channel
into five pions, thus its enhancement changes $\lambda_2$, the strength of two-pion
correlation functions at low pair momentum $K$~\cite{Vance:1998wd}.
This underlines the importance of understanding the effects that affect 
the measured $\lambda_2$ and its pair momentum dependence.

The possible presence of partially coherent pion production distorts the above picture~\cite{Sinyukov:1994en,Csorgo:1999sj,Csanad:2018vgk}.
It turns out however, that two-- and three-particle
Bose-Einstein correlation functions at zero relative momentum are in simple connection to the partially coherent
fraction ($p_c$) of the fireball~\cite{Csorgo:1999sj}:
\begin{align}
\lambda_2 &= f_c^2((1-p_c)^2+2p_c(1-p_c))\label{e:fcpc2}\\
\lambda_3 &= 2f_c^3((1-p_c)^3+3p_c(1-p_c)^2)+3f_c^2((1-p_c)^2+2p_c(1-p_c))\label{e:fcpc3}
\end{align}
 This means that a simultaneous measurement of $\lambda_2$ and $\lambda_3$ in two-- and
three-pion correlation functions offers the
possibility of investigation of coherent pion production, in addition to the resonance decay contribution.

\section{Strength of multi-particle Bose-Einstein correlations}
\label{s:lambda23}

Let us have a scenario with two point like sources, $a$ and $b$, at a distance of $R$, emitting particles with
wave functions $\Phi_a(r)$ and $\Phi_b(r)$. Let us furthermore have two detectors, $A$ and $B$, separated 
by $d$, and at an $L$ distance from the sources, with $d,R\ll L$. These detectors measure the total single particle
densities at their respective locations, $\Psi(r_A)$  and $\Psi(r_B)$. We may however also measure the two-particle
coinciding density $\Psi(r_A,r_B)$ i.e. the correlation function in detectors $A$ and $B$:
\begin{align}
C_{AB}=\frac{\langle |\Psi(r_A,r_B)|^2 \rangle}{\langle |\Phi(r_A)|^2 \rangle\langle |\Phi(r_B)|^2 \rangle} 
\end{align}
In this section we shall discuss how this correlation function looks like for thermally emitted particles, 
with a random field.


\subsection{Two-boson correlations with thermal emission}

Let the previously mentioned sources emit bosons with wavenumber $k$. Then the emitted matter-waves will have the form
\begin{align}
\Phi_a(r)&=\frac{1}{|r-r_a|}e^{ik|r-r_a|+i\phi_a},\\
\Phi_b(r)&=\frac{1}{|r-r_b|}e^{ik|r-r_b|+i\phi_b},
\end{align}
where $\phi_{a,b}$ are the (random) phases of the waves emitted from each point, 
while $k$ is the wavenumber of both waves. These particles are detected in detectors $A$ and $B$, 
where the two-particle wave-function is then
\begin{align}
&\nonumber \Psi(r_A,r_B)=\frac{1}{\sqrt{2}}\left(\Phi_a(r_A)\Phi_b(r_B)+\Phi_a(r_B)\Phi_b(r_A)\right)\nonumber \\&
=\frac{1}{\sqrt{2}}\left(\frac{1}{|r_A-r_a||r_B-r_b|}e^{ik|r_A-r_a|+ik|r_B-r_b|+i(\phi_a+\phi_b)}\right.\nonumber\\&
\left.+\frac{1}{|r_B-r_a||r_A-r_b|}e^{ik|r_B-r_a|+ik|r_A-r_b|+i(\phi_a+\phi_b)}\right)
\end{align}
The time-averaged single- and two-particle densities in detectors $A$ and $B$ are then
(based on the uniformly distributed random thermal phases)
in an approximation where $d,R\ll L$ (and defining $r_{aA}=|r_A-r_a|$ and similarly for $b$ and $B$):
\begin{align}
\langle |\Phi(r_{A,B})|^2 \rangle &= \frac{1}{L^2}\\
\langle |\Psi(r_A,r_B)|^2 \rangle&=\frac{1}{2L^4}
\left(2+e^{ik(r_{aA}+r_{bB}-r_{aB}-r_{bA})}+e^{-ik(r_{aA}+r_{bB}-r_{aB}-r_{bA})}\right)\nonumber\\
&=\frac{1}{L^4}\left(1+\cos\left(k\frac{Rd}{L}\right)\right)
\end{align}
as the phase average of factors like $e^{i(\phi_b-\phi_a)}$ is zero.
We may observe that $d/L$ is the angle between the two detectors, i.e. between the momenta
of the pair, thus $\frac{kd}{L}=\Delta k$, the momentum difference of the pair.
Then for the correlation function, one gets
\begin{align}
\frac{\langle |\Psi(r_A,r_B)|^2 \rangle}{\langle |\Phi(r_A)|^2 \rangle\langle |\Phi(r_A)|^2 \rangle} -1 = \cos\frac{kRd}{L} = \cos(R\Delta k)
\end{align}
At zero relative momentum, the correlation strength is then
\begin{align}
\left.\frac{\langle |\Psi(r_A,r_B)|^2 \rangle}{\langle |\Phi(r_A)|^2 \rangle\langle |\Phi(r_B)|^2 \rangle} \right|_{\Delta k=0} -1 = 1
\end{align}

\subsection{Effect of a random field on two-boson correlations}
If random phases have to be applied not just to the points of emittance, but also to the path,
then the wave functions are:
\begin{align}
\Phi_a(r)&=\frac{1}{|r-r_a|}e^{ik|r-r_a|+i\phi_a+i\phi(\textnormal{path to }r)}\\
\Phi_b(r)&=\frac{1}{|r-r_b|}e^{ik|r-r_b|+i\phi_b+i\phi(\textnormal{path to }r)}
\end{align}
The two-particle wave-function is then
\begin{align}
&\nonumber \Psi(r_A,r_B)=\frac{1}{\sqrt{2}}\left(\Phi_a(r_A)\Phi_b(r_B)+\Phi_a(r_B)\Phi_b(r_A)\right)\nonumber\\&=\frac{1}{\sqrt{2}}\left(\frac{1}{r_{aA}r_{bB}}e^{ikr_{aA}+ikr_{bB}+i(\phi_a+\phi_b)+i(\phi_{aA}+\phi_{bB})}\right.\nonumber\\&
\left.+\frac{1}{r_{bA}r_{aB}}e^{ikr_{aB}+ikr_{bA}+i(\phi_a+\phi_b)+i(\phi_{aB}+\phi_{bA})}\right)
\end{align}
One can again form the phase- or time-average of the single- and two-particle density in detectors $A$ and $B$,
in an approximation where $d,R\ll L$:
\begin{align}
&\langle |\Phi(r_{A,B})|^2 \rangle = \frac{1}{L^2}\\&
\nonumber\langle |\Psi(r_A,r_B)|^2 \rangle=\frac{1}{2L^4}
\left(2+e^{ik(r_{aA}+r_{bB}-r_{aB}-r_{bA})+i(\phi_{aA}+\phi_{bB}-\phi_{aB}-\phi_{bA})}\right.\nonumber\\&
\left.+e^{-ik(r_{aA}+r_{bB}-r_{aB}-r_{bA})-i(\phi_{aA}+\phi_{bB}-\phi_{aB}-\phi_{bA})}\right)\nonumber\\&
=\frac{1}{L^4}\left(1+\cos\left(k\frac{Rd}{L}+\phi\right)\right)
\end{align}
where $\phi$ is the total phase picked up through the random route. Normalized by the single-particle densities, one gets
\begin{align}
\frac{\langle |\Psi(r_A,r_B)|^2 \rangle}{\langle |\Phi(r_A)|^2 \rangle\langle |\Phi(r_B)|^2 \rangle} -1 = \cos\left(\frac{kRd}{L}+\phi\right) = 1+\cos(R\Delta k+\phi)
\end{align}
and thus averaged over a Gaussian distribution of $\phi$ values, one gets
\begin{align}
\frac{\langle |\Psi(r_A,r_B)|^2 \rangle}{\langle |\Phi(r_A)|^2 \rangle\langle |\Phi(r_B)|^2 \rangle} -1 = \cos(R\Delta k)e^{-\frac{\sigma^2}{2}}
\end{align}
At zero relative momentum, the correlation strength is then
\begin{align}
\left.\frac{\langle |\Psi(r_A,r_B)|^2 \rangle}{\langle |\Phi(r_A)|^2 \rangle\langle |\Phi(r_B)|^2 \rangle} \right|_{\Delta k=0} -1
 = e^{-\frac{\sigma^2}{2}}
\end{align}

\subsection{Three-boson correlations with thermal emission}
Let us now again turn to three-particle correlations. In this case, the three-particle symmetrized wave function is
\begin{align}
&\nonumber \Psi(r_A,r_B,r_C)=\frac{1}{\sqrt{6}}(
\Phi_a(r_A)\Phi_b(r_B)\Phi_c(r_C)+
\Phi_a(r_B)\Phi_b(r_C)\Phi_c(r_A)+\\&
\Phi_a(r_C)\Phi_b(r_A)\Phi_c(r_B)+
\Phi_a(r_C)\Phi_b(r_B)\Phi_c(r_A)+
\Phi_a(r_A)\Phi_b(r_C)\Phi_c(r_B)\nonumber \\&+
\Phi_a(r_B)\Phi_b(r_A)\Phi_c(r_C))
=\frac{1}{\sqrt{6}L^3}e^{i(\phi_a+\phi_b+\phi_c)}(
e^{ik(r_{aA}+r_{bB}+r_{cC})}\nonumber \\&+
e^{ik(r_{aB}+r_{bC}+r_{cA})}+
e^{ik(r_{aC}+r_{bA}+r_{cB})}+
e^{ik(r_{aC}+r_{bB}+r_{cA})}\nonumber\\&+
e^{ik(r_{aA}+r_{bC}+r_{cB})}+
e^{ik(r_{aB}+r_{bA}+r_{cC})})
\end{align}
where the factor $\exp i(\phi_a+\phi_b+\phi_c)$ is contained in each term. Then the phase-averaged three-particle density is:
\begin{align}
\langle |\Psi(r_A,r_B,r_C)|^2 \rangle&=\frac{1}{6L^6}
\left(6+(\textnormal{30 other cross-terms})\right)
\end{align}
we refrain here from writing all of them out, but point out that  at zero relative momenta all terms become unity at, thus
\begin{align}
\left.\frac{\langle |\Psi(r_A,r_B,r_C)|^2 \rangle}{\langle |\Phi(r_A)|^2 \rangle\langle |\Phi(r_B)|^2 \rangle\rangle\langle |\Phi(r_C)|^2 \rangle}\right|_{\Delta k=0}-1 =5
\end{align}
Note that here $k\Delta r$ type of terms were converted to $R\Delta k$. This is possible due to geometrical symmetry, as discussed
in the first sections.

\subsection{Effect of a random field on three-boson correlations}
If there are random fields picked up in different paths, then these enter in the tree-particle wave function as
\begin{align}
&\nonumber \Psi(r_A,r_B,r_C)=\frac{1}{\sqrt{6}L^3}e^{i(\phi_a+\phi_b+\phi_c)}\\\nonumber
&\times\left(
e^{ik(r_{aA}+r_{bB}+r_{cC})+i(\phi_{aA}+\phi_{bB}+\phi_{cC})}+
e^{ik(r_{aB}+r_{bC}+r_{cA})+i(\phi_{aB}+\phi_{bC}+\phi_{cA})}\right.\nonumber\\&+
e^{ik(r_{aC}+r_{bA}+r_{cB})+i(\phi_{aC}+\phi_{bA}+\phi_{cB})}+
e^{ik(r_{aC}+r_{bB}+r_{cA})+i(\phi_{aC}+\phi_{bB}+\phi_{cA})}\nonumber\\&+\left.
e^{ik(r_{aA}+r_{bC}+r_{cB})+i(\phi_{aA}+\phi_{bC}+\phi_{cB})}+
e^{ik(r_{aB}+r_{bA}+r_{cC})+i(\phi_{aB}+\phi_{bA}+\phi_{cC})}\right)
\end{align}
and from this, we have to calculate $\langle |\Psi(r_A,r_B,r_C)|^2 \rangle$ again. We can observe that in this case there will be
three type of terms:
\begin{itemize}
\item 6 terms like $|e^{ik(r_{aA}+r_{bB}+r_{cC})+i(\phi_{aA}+\phi_{bB}+\phi_{cC})}|^2=1$
\item 6 terms where e.g. $r_{aA}+r_{bB}+r_{cC}$ meets $r_{aB}+r_{bA}+r_{cC}$, and in this case,
the result is a pair-correlation type of term, i.e. $e^{ik(r_{aA}+r_{bB}-r_{aB}-r_{bA})+i(\phi_{aA}+\phi_{bB}-\phi_{aB}-\phi_{bA})}$;
we get all 3 such terms and their complex conjugates as well
\item 12 ``almost'' pair-correlation like terms, where e.g. $r_{aB}+r_{bC}+r_{cA}$ that meets $r_{aB}+r_{bA}+r_{cC}$; in this case, the result is $e^{ik(r_{bC}+r_{cA}-r_{bA}-r_{bA})+i(\phi_{bC}+\phi_{cA}-\phi_{aB}-\phi_{cC})}$; these don't represent closed loops, but contain only four paths.
\item 12 terms containing nine paths i.e. nine $i\phi_{xX}$ like terms in the exponent
\end{itemize}
all these get different weights when averaged on all the $\phi_{xX}$ phases.

Now, let us try to estimate the value of the correlation function at zero relative momenta. Previously, we introduced a single phase $\phi$ as a sum of four $\phi_{xX}$ like phases, and this was supposed to have a Gaussian distribution of $\exp -\phi^2/(2\sigma^2)$. Based on the summing of
random variables, this means that a single $\phi_{xX}$ like phase has to have the distribution of $\exp -\phi^2/(2(2\sigma)^2)$, i.e. a double
width of $2\sigma$. Henceforth the terms that contain a sum of four paths and four phases in the exponent will again have a distribution of 
$\exp -\phi^2/(2\sigma^2)$, while the terms with nine paths will have a multiplier of $\exp -\phi^2/(2(2\sigma/3)^2)$. In the end, the result will be
\begin{align}
&\nonumber \left.\frac{\langle |\Psi(r_A,r_B,r_C)|^2 \rangle}{\langle |\Phi(r_A)|^2 \rangle\langle |\Phi(r_B)|^2 \rangle\rangle\langle |\Phi(r_C)|^2 \rangle}
\right|_{\Delta k=0} -1 =  \nonumber \\&
\frac{1}{6}\left( 6 + 18 e^{-\frac{\sigma^2}{2}} + 12 e^{-\frac{(2\sigma/3)^2}{2}} \right)-1 
= 3e^{-\frac{\sigma^2}{2}}+2e^{-\frac{2\sigma^2}{9}}
\end{align}


\section{Toy model and calculations}\label{s:toymodel}

After rehadronization in heavy-ion collisions, hundreds of charged particles are produced. When measuring the correlation functions, we take into account that the produced hadrons create a strong electromagnetic field around trajectories of the investigated pairs of identical pions. Although this may be seen as an Aharonov-Bohm effect, a more straightforward explanation would be that the phase along the pair's closed path is altered when  one of the particles' paths is altered by a phase, as compared to the interaction-free case, when the path is a straight line, and momentum also does not change. This additional phase shift for an infinitesimal path element $dx$ can be expressed as $k\cdot dx$, where $k=p/\hbar$ is the momentum (or wavenumber) of the particle at that point. The alteration of the particles' flight time reaching the detector can be connected to the phase shift of the particles, as we discuss below.

The model we set up describes the path of the particles from the collision point to the detector based on equations~(\ref{eq:force})-(\ref{eq:velocity}). We calculate the phase shift by taking into account the Aharonov-Bohm-like effect in form of the Coulomb interaction between the probe particles with charged particles of the cloud. This cannot be solved analytically, so we apply a numerical calculation, as detailed below. The $N_{ch}$ produced charged particles are normally distributed (Gaussian distribution) with zero total charge and with fireball radius $R$, undergoing 3D Hubble flow which describes the time evolution of the produced hadron gas of charged particles after the freeze-out process. Thus for the location vector ${\boldsymbol r}$ of the fireball particles, we follow a simple time-dependence determined by $\dot {\boldsymbol r}(t) = H(t)\cdot {\boldsymbol r}(t)$, where $H(t)=1/t$ is the Hubble-constant (constant in space, but inversely proportional to time). We then follow the movement of a charged probe particle, affected by all the particles of the charged cloud (without backreaction). We prepared the simulation code in C++ using the Euler method with time iteration, tracking the movement of a pion with mass $m \approx 139.57$ MeV (utilizing $c=1$ units), from an initial location characterized by a distance $d$ from the origin.
There are several simulated scenarios with initial momentum in the range of $10 - 300$ MeV (again in $c=1$ units), and tracked distances from $100$ fm to $50000$ fm (when the simulation is terminated and accumulated time-difference and the resulting phase-shift is calculated). Then the dynamical equations of the system are the following:
\begin{align} 
    \frac{d\boldsymbol{p}}{dt} &= \hbar c \alpha  \sum_{j=1}^{N_{ch}}\frac{q(\boldsymbol{r}_j - \boldsymbol{X})}{r^3}, \label{eq:force}\\
    \frac{d\boldsymbol{X}}{dt} &= \boldsymbol{V} = \frac{\boldsymbol{p}}{m\gamma}\label{eq:velocity}
\end{align}
where 
\begin{itemize}
    \item $\boldsymbol{X}$ is the position of the probe particle,
    \item $\boldsymbol{r}_j$ denotes the position of the charged particles of the cloud, updated in time via the above mentoned equation ($\dot {\boldsymbol r}_j = H\cdot \boldsymbol{r}_j$),
    \item $q = \pm 1$ is the pion charge (in units of the elementary charge),
    \item $\hbar c \alpha \approx \frac{197.326}{137.036}$ MeV fm,
    \item $r = \sqrt{(\boldsymbol{r_{j}} - \boldsymbol{X})^2}$,
    \item $\gamma = \sqrt{1 - \frac{\boldsymbol{p}^2}{m^2}}$ is the Lorentz-factor,
     \item $\boldsymbol{p}$ is the momentum of the pion, 
     \item $m \approx 139.57$ MeV is the pion mass(using the natural units c=1),
     \item and the summation $j=1\dots N_{ch}$ goes over cloud particles.
\end{itemize}

\begin{figure}
    \centering
    \includegraphics[width=\textwidth]{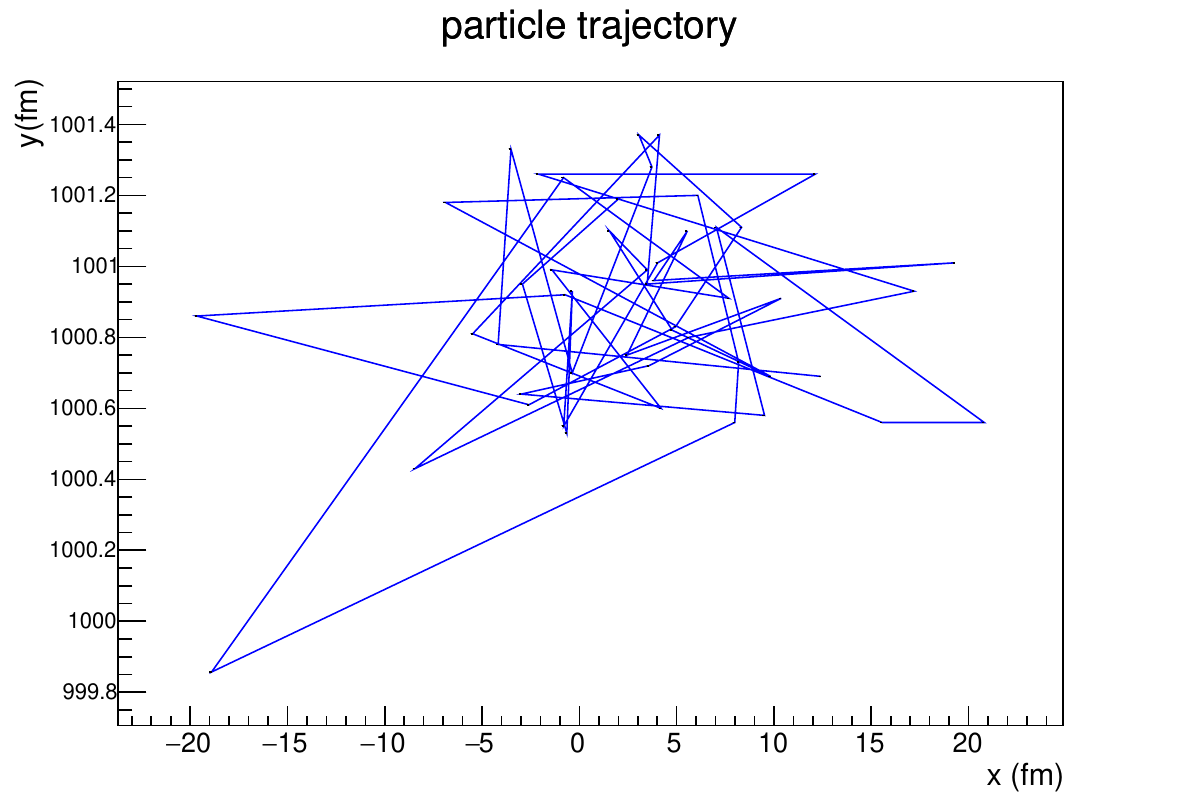}
    \caption{Example particle trajectory of a particle with initial momentum vector (0,0,$p_{z}$) with $p_{z}=300$ MeV.}
    \label{fig:particle_trjectory}
\end{figure}

From the numerical solution of that model, we get an example of a particle trajectory inside the potential of the charged cloud when considering the final state effects as an Aharonov-Bohm effect is shown in fig.(\ref{fig:particle_trjectory}) with initial transverse momentum $p_{z} = 300$ MeV.

In this model, we can measure the $t_{\rm TOF}(d)$ time that is needed by the particle to travel a distance $d$.
If $N_{\rm ch}=0$ (free case), then (in $c=1$ units)
\begin{align}
t_{\rm TOF}^{(0)}(d) = d\sqrt{1+\frac{m^2}{p^2}}.
\end{align}
If $N_{\rm ch}>0$, the particle is (relativistically) accelerated and decelerated by the same- and opposite-sign charged particles of the cloud,
respectively. We studied fluctuating charge clouds (with variable $N_{\rm ch}$, $R$ and $d$ values), from which Gaussian distributions of $\Delta t = t_{\rm TOF}(d)-t_{\rm TOF}^{(0)}(d)$ emerged, with a width $\sigma_t$
depending on initial momentum $p_{t}$of the probe particle.

\section{Results}\label{s:results}

\begin{figure}
    \centering
    \includegraphics[width=\textwidth]{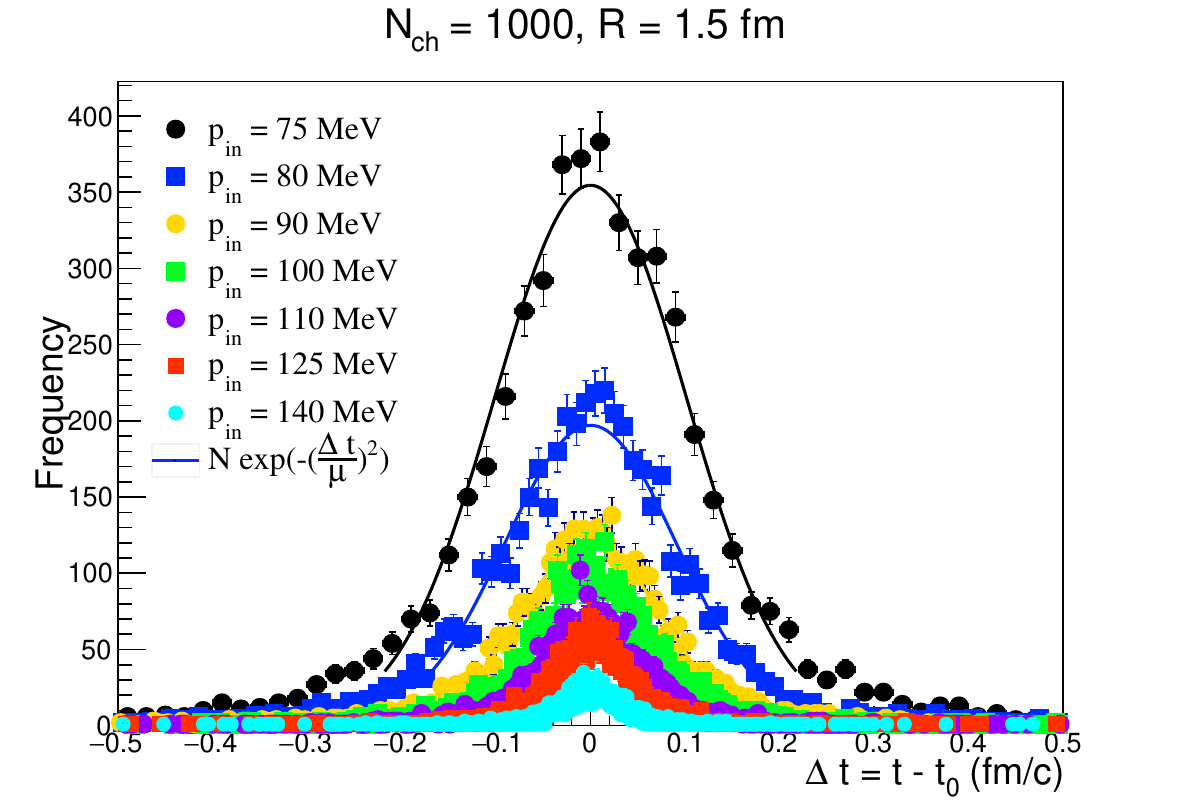}
    \caption{Time shift distributions from the simulation for different initial momentum values.}
    \label{fig:time-sh dist}
\end{figure}

From our simulations, we calculate the $\Delta t$ difference between the expected arrival time after modification and the arrival time in the free case (ignoring the final state interaction) which appears to follow a normal distribution, as shown in fig. \ref{fig:time-sh dist}. The results of the Gaussian fits to a few example distributions are shown in table \ref{tab:fitresults}.

\begin{table}
    \centering  
    \begin{tabular}{|l|c|c|c|r|}
    \hline
        $p_{\rm init}$ [MeV]  & $N$ & $\mu$ [fm$/c$] & $\sigma_{t}$ [fm$/c$] & $\chi^2/ndf$ \\ \hline
        75 & $354.5 \pm 7.2 $& $0.144\pm 0.0023$ & $0.102 \pm 0.0015$ & $38.56/20$ \\ 
        \hline
        80 & $197.1 \pm 4.19$ & $0.129\pm 0.0024$& $0.091 \pm 0.0019$ & $59/34$ \\
        \hline
        90 & $121.0 \pm 2.5$ & $0.104 \pm 0.0018$ & $0.074 \pm 0.0008 $ & $63.02/58$ \\
        \hline 
        100 & $94.0 \pm 2.0$ & $0.089\pm 0.0016$ & $0.063 \pm 0.0011 $ & $113.9/76$ \\
        \hline 
        110 & $ 68.2\pm 1.4$ & $0.0733 \pm 0.0014$ & $0.052 \pm 0.0010$ & $98.86/98$ \\
        \hline
        125 & $54.2\pm 1.1$ & $0.058 \pm 0.00085$ & $0.041 \pm 0.00061 $ & $188.6/158$ \\
        \hline 
        140 & $21.0 \pm 0.44$ & $0.0446 \pm 0.00067$ & $0.0315 \pm 0.0005 $ & $514.5/435$ \\
        \hline 
    \end{tabular}
    \vspace{5pt}
    \caption{Fitting results (c.f. figure~\ref{fig:time-sh dist}) Gaussian fit function for different initial momenta with $R=1.5$ fm and $N_{ch} = 1000$ charged paricle and $d = 1600$ fm. Probabilities ($p$-values) of all fits are above 0.1\%. Here $N$ is a normalization parameter and $\mu$ is a width parameter, as indicated in the legend of figure~\ref{fig:time-sh dist}. From this, the standard deviation variable $\sigma_t$ results by a division by $\sqrt{2}$.}
    \label{tab:fitresults}
\end{table}

\begin{figure}
\centering
    \includegraphics[width=\textwidth]{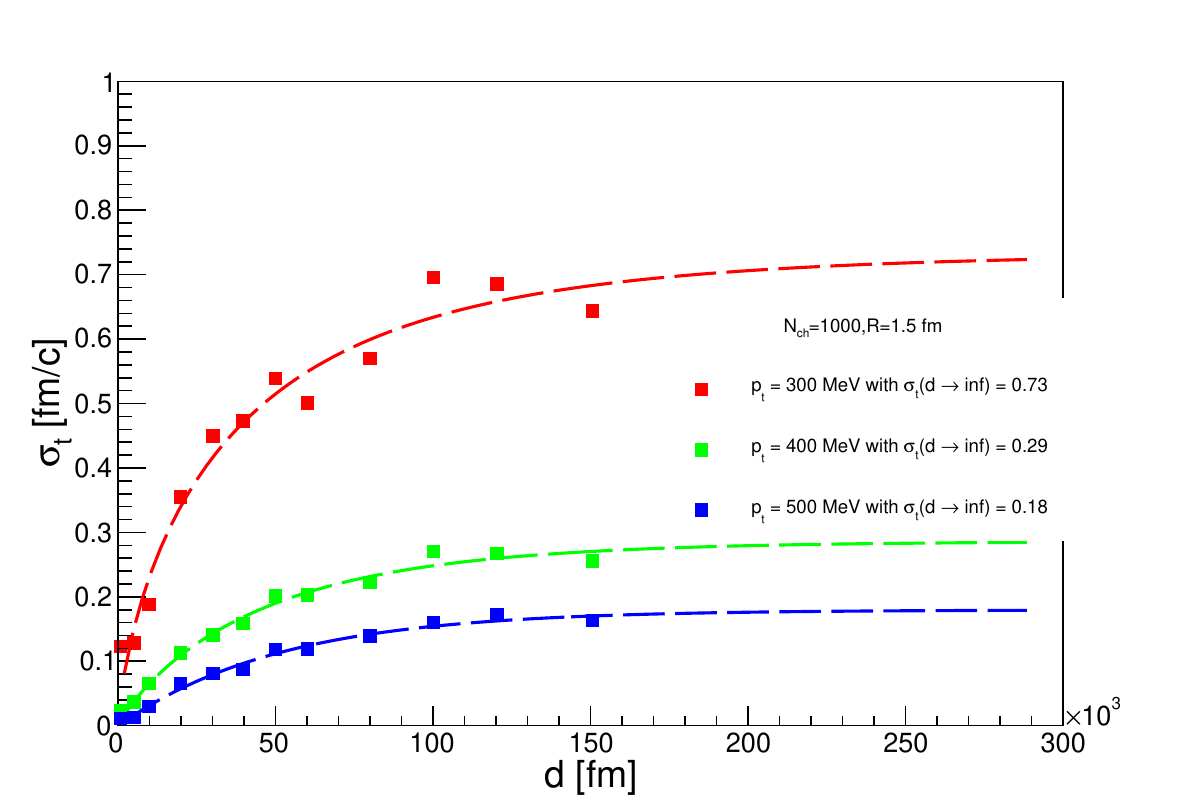}
    \caption{ The traveled distance by the investigated correlated particles with the phase-shift distribution $\sigma_0$, for $N_{\rm ch}=1000$, $R=1.5$ fm.}
    \label{f:extrapolation_distances}
\end{figure}

\begin{figure}
    \centering
    \includegraphics[width=\textwidth]{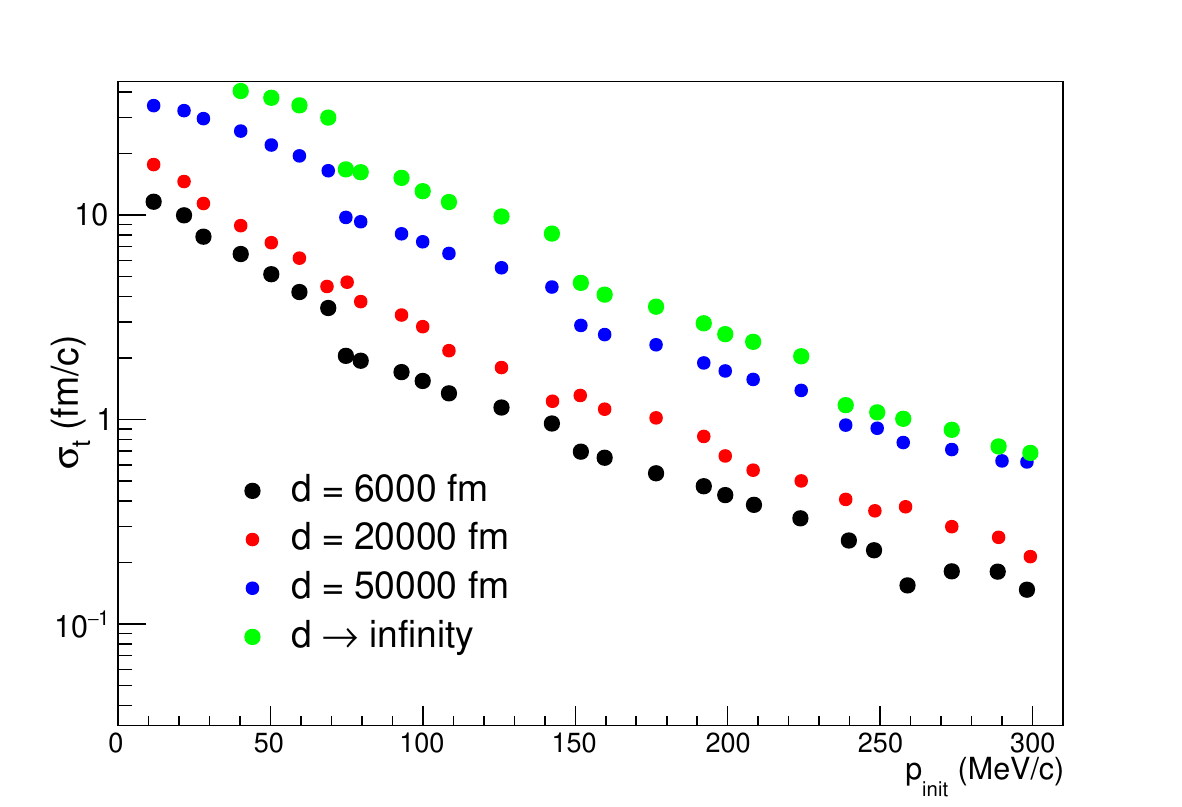}
    \caption{The dependency of the gaussian width and then the phase shift on the initial momentum of the correlated particles.}
    \label{fig:sigma_t with p}
\end{figure}

While our calculations always run until a finite distance $d$ is reached, we are interested in the limiting case of $d\rightarrow \infty$, or a distance corresponding to actual detector sizes, practially infinity, compared to our femtometer (or up to a few hundred picometer) distances. Thus we investigate the distance-dependence of the phase-shift width, and fit it with a convergent function of the following shape:
\begin{align}
    \sigma_t(d) = A (1 - e^{-Bd^{C}})
\end{align}
where $A$, $B$, and $C$ are auxiliary parameters, and then $\sigma_t(d\rightarrow \infty) = A$. Several such example fits are shown in fig. \ref{f:extrapolation_distances}. In the end we use this extrapolated width, as a function of particle momentum $p$, corresponding to transverse momentum $p_t$ at or near midrapidity. The results for this extrapolated $\sigma_t$, as a function of initial particle momentum, can be seen in Fig. \ref{fig:sigma_t with p}. 
It is clear that when the initial transverse momentum of the probe particle increases, the interactions with the cloud charge and Aharanov-Bohm effect decrease, and then the time-shift width decreases.

The time shift $\Delta t$ is then connected to the phase-shift through velocity $v$ and wavenumber $k$ (assuming negligible change of momentum) as
\begin{align}
\phi = k\Delta x =  \frac{p}{\hbar}v \Delta t =  \frac{p^2}{\hbar\sqrt{m^2+p^2}}\Delta t.
\end{align}
Hence the width of the time-shift distribution $\sigma_t$ is a good quantifier for $\sigma$ the width of the phase-shift distribution:
\begin{align} \label{sigmap}
\sigma = \frac{p^2}{\hbar\sqrt{m^2+p^2}}\sigma_t.
\end{align}

\begin{figure}
    \centering
    \includegraphics[width=\textwidth, height=7.5cm]{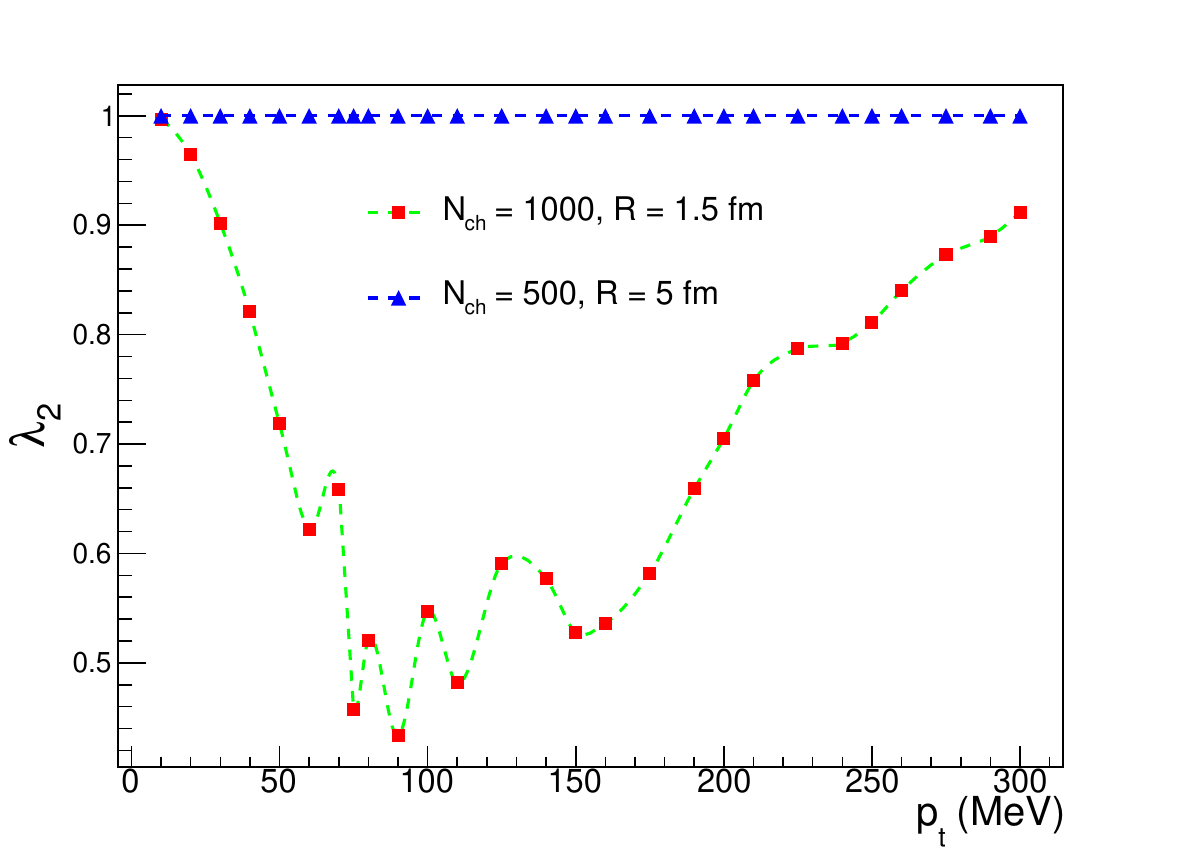}
    \caption{The intercept parameter $\lambda_2$ as a function of the initial transverse momentum of the probe particle, for the two scenarios: $N_{\rm ch}= 500$, $R = 5$ fm and $N_{\rm ch}=1000$, $R = 1.5$ fm.}
    \label{fig:Intercept parameters}
\end{figure}

In this analysis, we investigated two main scenarios. The first one when $N_{\rm ch} = 500$ particle with $R_{\rm fireball} = 5$ (fm) and the second one when $N_{\rm ch} = 1000$ particle with $R_{\rm fireball} = 1.5$ fm, where $N_{\rm ch}$ is the number of charged particles in the charged cloud of pions, and $R_{\rm fireball}$ is the fireball radius of the cloud.
From the second scenario ($N_{\rm ch} = 1000$ and $R = 1.5$ fm) the intercept parameters of Bose-Einstein correlations, as shown in fig.~\ref{fig:Intercept parameters}, show a significant decrease for the final state effects.

\begin{figure}
    \centering
    \includegraphics[width=\linewidth, height=7.5cm]{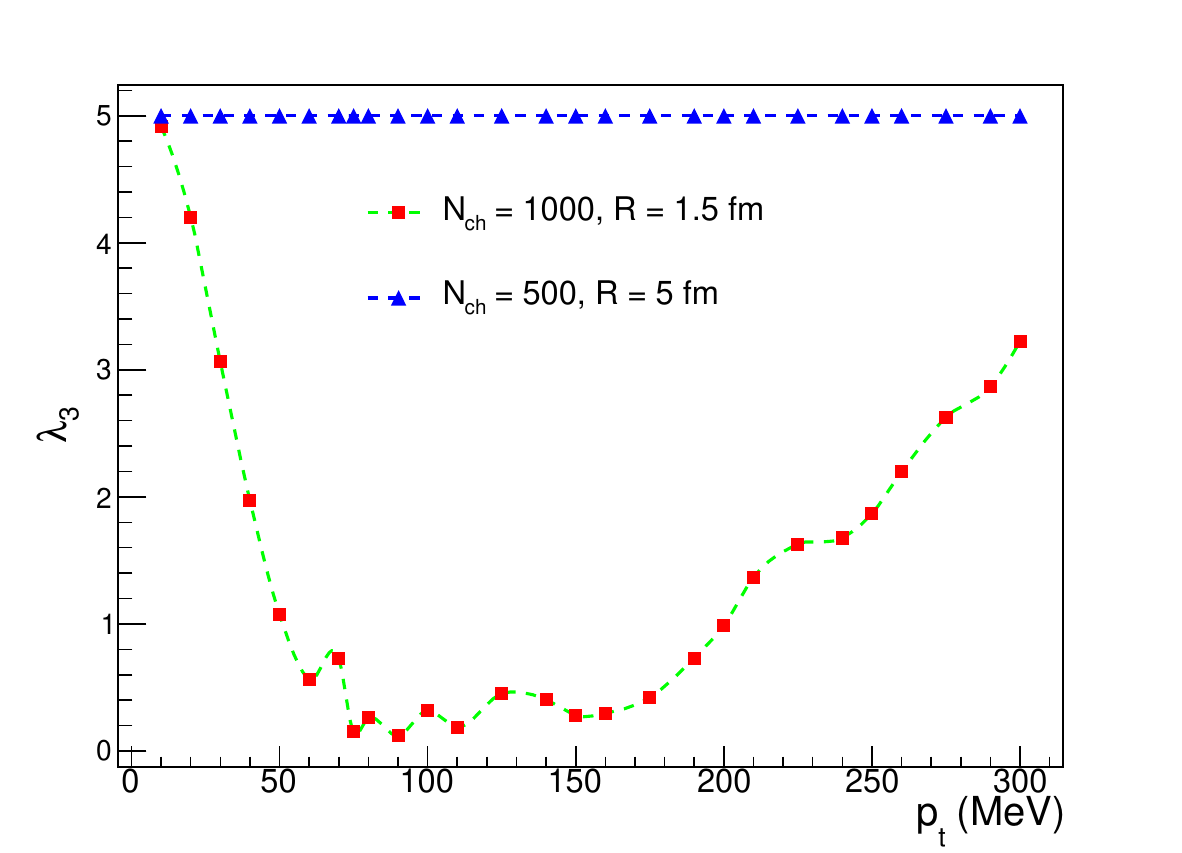}
    \caption{The intercept parameter $\lambda_3$ as a function of the initial transverse momentum of the probe particle, for the two scenarios: $N_{\rm ch}= 500$, $R = 5$ fm and $N_{\rm ch}=1000$, $R = 1.5$ fm.}
    \label{fig:lamda_3}
\end{figure}

\begin{figure}
    \centering
    \includegraphics[width=\textwidth, height=7.5cm]{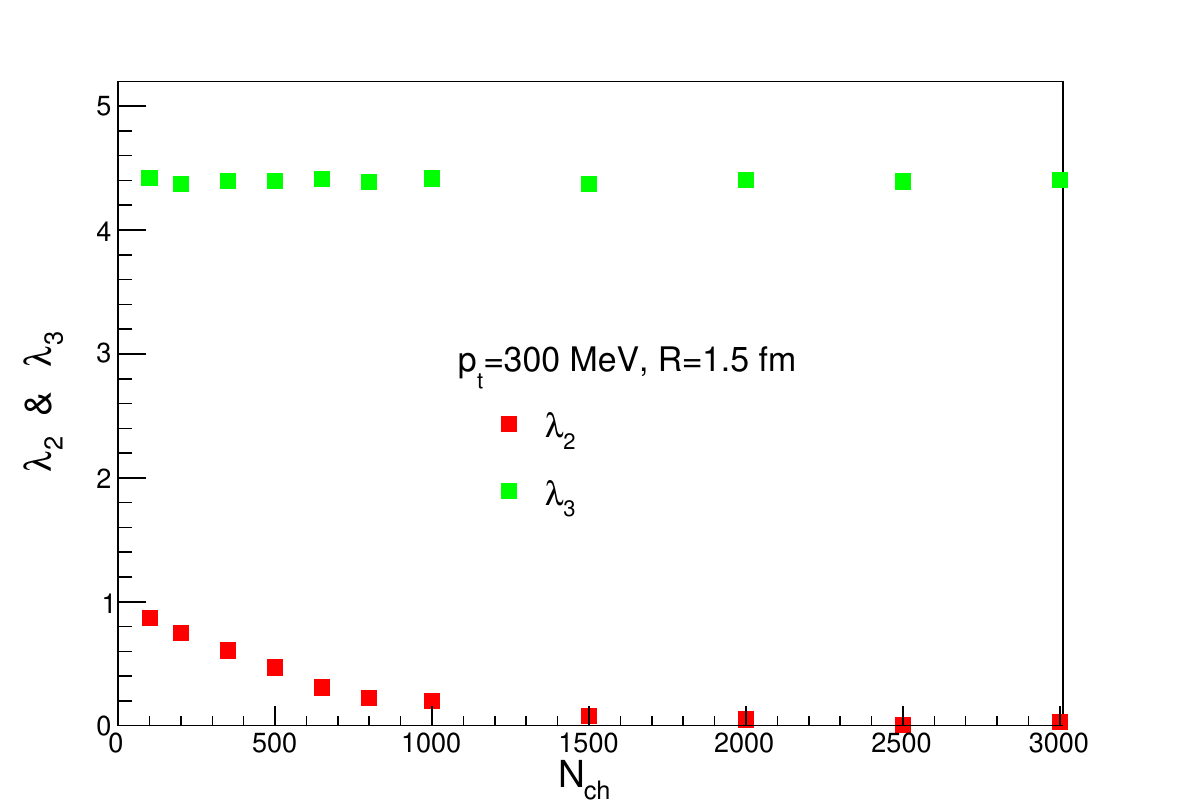}
    \caption{Correlation strength parameters versus charged particle multiplicity.}
    \label{fig:multiplicity}
\end{figure}

\begin{figure}
    \centering
    \includegraphics[width=\linewidth, height=7.5cm]{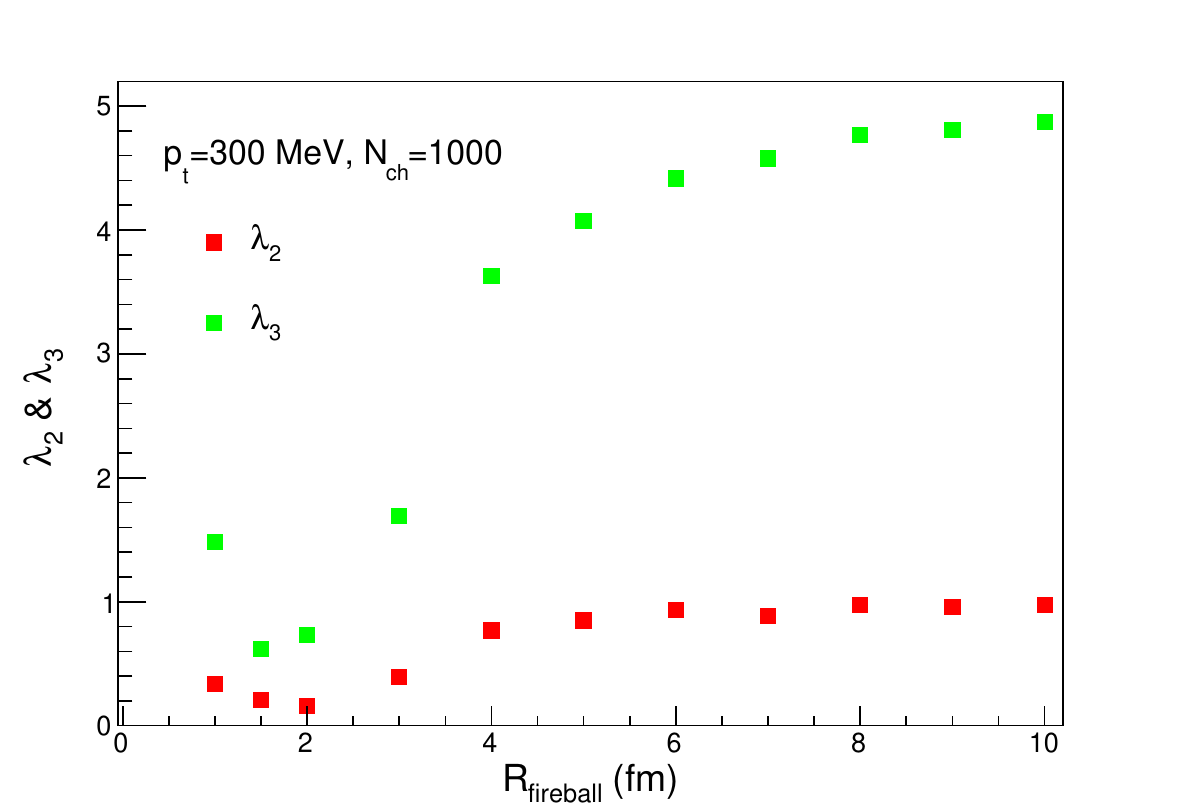}
    \caption{Correlation strength parameters versus fireball radius.}
    \label{fig:firba}
\end{figure}

\begin{figure}
    \centering
    \includegraphics[width=\textwidth, height=7.5cm]{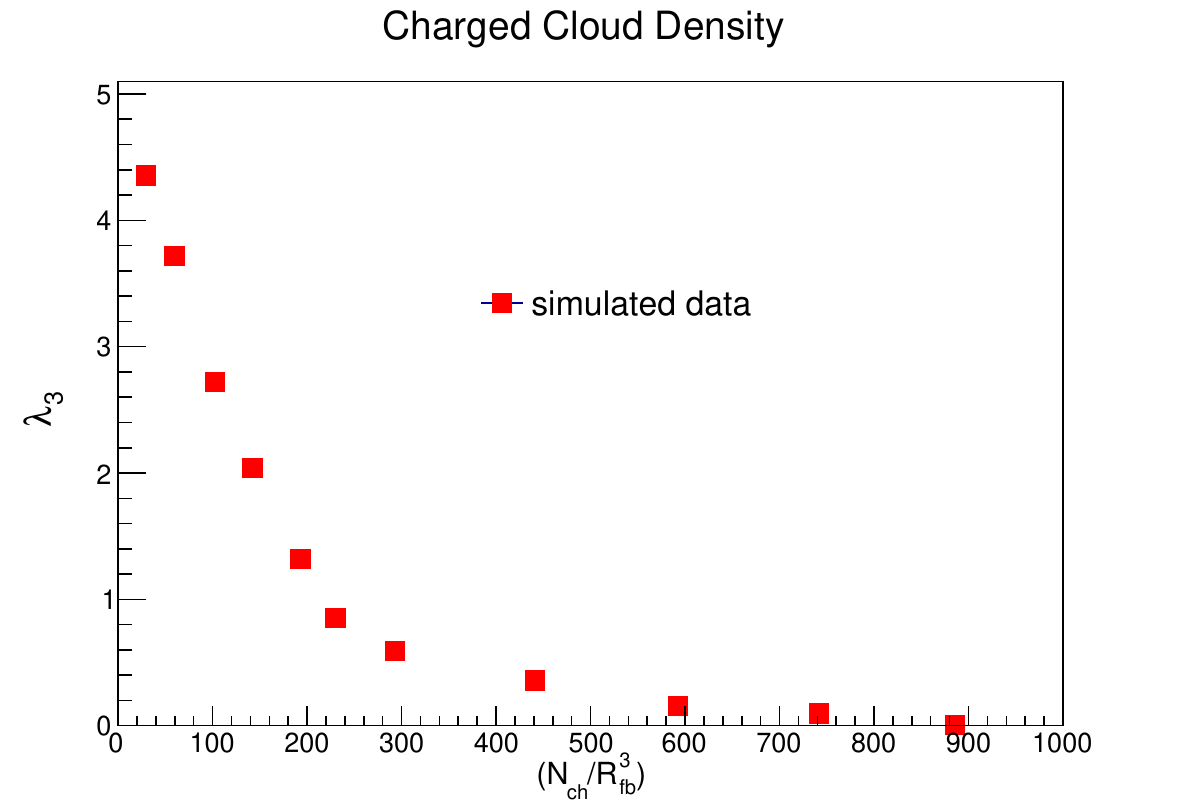}
    \caption{Intercept parameter $\lambda_3$ versus a charged particle density proxy.}
    \label{fig:Multiplicity and density}
\end{figure}

If the standard deviation of the Gaussian distribution in time of flight shift $\sigma_{t}$ can be estimated using ROOT tool, then $\sigma(p)$ may be calculated as given in equation (\ref{sigmap}). We may substitute this into
\begin{align}
\lambda_2 &= e^{-2\sigma^2},\textnormal{ and }\\
\lambda_3 &= 3 e^{-2 \sigma^2} + 2 e^{-3\sigma^2}.
\end{align}
The resulting parameters are shown in figs.~\ref{fig:Intercept parameters} and \ref{fig:lamda_3}. 
In addition, the dependence on the fireball radius charged particle multiplicity is clear from figs.~\ref{fig:multiplicity} and \ref{fig:firba}. Finally, $\lambda_3$ is shown as a function of a density profy, $N_{\rm ch}/R_{\rm fireball}^3$, in fig.~\ref{fig:Multiplicity and density}, and is found to systematically decrease with it.

\section{Conclusion}
The study investigates the space-time structure and inner dynamics of ultra-relativistic collisions like heavy ion collisions and pp collisions with high multiplicity after the freeze-out of very hot and dense matter (Quark-Gluon Plasma) using a simulation via a toy model. Changes in phases due to the Aharonov-Bohm effect, effectuated via the Coulomb interaction, may cause distortion in quantum-statistical correlations. The simulation was established from the numerical solution of the movement of particles to measure the modification of the strenght of correlation functions, via the time-shift of the arrival time. The study found that the change in phases is clear at low momenta of the investigated particle and decreases at high momenta. The study also found that this correlation strenght change depends on the charged particle density, and that for high densities this additional effect may be important to consider. This procedure would be rather resource-comsuning for state-of-the art Monte Carlo simulations of high-energy heavy-ion collisions, due to the large range of the Coulomb interaction, as well as the large timescale (hundreds of thousands of fm$/c$ units) required for the effect to be formed. Nevertheless, when calculating momentum correlations, it can be taken into account via an afterburner, operating along the procedure outlined in this paper.

\bibliographystyle{unsrt}

\section*{Acknowledgments}
Hemida H. Mohammed would like to thank Dr. Ahmad Lotfy for his helpful discussions. This work was supported by the Hungarian NFKIH grant K-138136. M.A. Mahmoud and Hemida Hamed Mohammed were supported by Science, Technology, and Innovation Funding Authority (STDF)  GE-SEED call grant ID: 43971.

\bibliography{sample}

\end{document}